

\input harvmac
\input epsf
\def\NP#1{Nucl.\ Phys. {\bf B{#1}}}
\def\PL#1{Phys.\ Lett. {\bf B{#1}}}
\def\PREP#1{Phys.\ Rep. {\bf {#1}}}

\Title{\vbox{\baselineskip12pt\hbox{KYUSHU-HET-7}\hbox{September, 1993}}}
{\vbox{\centerline{On a mean field approximation}
       \vskip3pt
       \centerline{for Higgs-Yukawa systems}}}

\centerline{Sergei V. Zenkin \footnote{$^*$}{Permanent address: Institute for
Nuclear Research of the Russian Academy of Sciences, Moscow 117312, Russia.
E-mail addresses: zenkin@jpnyitp.bitnet, h79104a@kyu-cc.cc.kyushu-u.ac.jp}}
\bigskip \centerline{Department of Physics, Kyushu University 33}
\centerline{Fukuoka 812, Japan}
\vskip 1.5cm

We discuss the phase structure of a lattice Higgs-Yukawa system in the
variational mean field approximation with contributions of fermionic
determinant
being calculated in a ladder approximation. In particular, we demonstrate
that
in this approximation the ferrimagnetic phase in the $Z_2$ model with naive
fermions can appear as an artifact of a finite lattice and that the phase
diagram for this model on infinite lattice changes qualitatively at
space-time
dimension $D = 4$ compared with those at $D > 4$.
\Date{}

\newsec{Introduction}

Although mean field method for lattice systems including fermions loses
considerably its simplicity and requires further approximations it is still
useful to get some idea of the phase structure of the systems and to
orientate
Monte Carlo simulations towards investigating the most interesting points. In
this paper we make an improvement in the approximations within the
variational
mean field approximation for $Z_2$ Higgs-Yukawa systems by summing up a
ladder
type contributions to fermionic determinant, including those of the next
order
in inverse space-time dimension $1/D$. This enables us to observe two new
points. As the first one we demonstrate that within our approximation the
ferrimagnetic phase in the simplest Higgs-Yukawa model with naive fermions
can
arise as a finite lattice artifact. The second point is that the value $D =
4$
turns out in a sense to be critical, as the domain of paramagnetic phase just
at $D = 4$ becomes disconnected, being connected at $D > 4$.

The paper is organized as follows. The system under consideration is defined
in
Sect. 2. In Sect. 3 we describe the method and approximations. Results are
discussed in Sect.4.

\newsec{The model}

The system is defined on a hyper cubic $D$-dimensional ($D$ is even) lattice
$\Lambda$ with sites numbered by $n = (n_1, ..., n_D)$, $-N/2+1 \leq n_{\mu}
\leq N/2$ ($N$ is even) and with lattice spacing $a = 1$; $\hat{\mu}$ is the
unit vector along the lattice link in the positive $\mu$-direction. Dynamical
variables of the model are the fermion $2^{D/2}$-component fields $\psi_n$,
$\psibar_n$, and scalar field $\phi_n \in Z_2$ (i.e. $\phi_n = \pm 1$). We
imply
antiperiodic boundary conditions for the fermion and periodic for the scalar
fields.

The model is defined by functional integral
\eqn\sys{Z[J] = \sum_{\phi_n \in Z_2} \int \prod_{n \in \Lambda} d\psi_n
d\psibar_n  \e{ -A[\phi, \psi, \psibar] + \sum_n J_n \phi_n}}
with the action
\eqn\act{A[\phi, \psi, \psibar] = - 2 \kappa \sum_{n, \mu} \phi_n
\phi_{n+\hat{\mu}} + \sum_{m, n} \psibar_m (\dsl_{mn} +  y \phi_m
\delta_{mn})\psi_n,} where
\eqn\d{\dsl_{mn} = \sum_{\mu} \gamma_{\mu} {1\over 2} (\delta_{m + \hat{\mu}
\;
n}  - \delta_{m - \hat{\mu} \; n}) = N^{-D} \sum_{p, \mu} \e{i p (m - n)}
i \, \gamma_{\mu} \, L_{\mu}(p),}
$\kappa \in (-\infty, \infty)$ is the hoping parameter, $y \geq 0$ is
the Yukawa coupling; we use the Hermitean $\gamma$-matrices: $[\gamma_{\mu},
\gamma_{\nu}]_{+} = 2 \delta_{\mu \nu}$; $L_{\mu}(p) = \sin p_{\mu}$,
$p_{\mu}  = (2 \pi / N) (k_{\mu} - 1/2)$,  $-N/2+1 \leq k_{\mu}\leq N/2$,
so that $p_{\mu} \in (-\pi/2, \pi/2)$. Operator $\dsl$ satisfies
\eqn\pr{\dsl_{mn} = -\dsl_{nm}.} In the limit of $N
\rightarrow \infty$ the sum $N^{-D} \sum_p$ defines the  integral
$\int^{\pi/2}_{-\pi/2} d^D p / (2 \pi)^D$.

The action \act\ is invariant under $Z_2$ global chiral transformations
\eqn\trans{\phi_n \rightarrow -\phi_n,\; \; \;
\psi_n \rightarrow (-P_L + P_R) \psi_n,\; \; \; \psibar_n \rightarrow
\psibar_n
(-P_R + P_L),}
where $P_{L,R} = (1 \pm \gamma_{D+1})/2$ are chiral projecting operators.

\newsec{The method and approximations}

To analyze the phase structure of the model we use the variational mean
field approximation \ref\DZ{J.-M. Drouffe, J.-B. Zuber, \PREP{102}
(1983) 1} (see also \ref\STs{M. A. Stephanov, M. M. Tsypin, \PL{236}
(1990) 344;
Sov. Phys. JETP 70 (1990) 228})  which becomes applicable to \sys\ after
integrating out the fermions \eqn\Z{ Z[J] = \e{- W[J]}  = \sum_{\phi_n \in
Z_2}
\e{ 2 \kappa \sum_{n, \mu} \phi_n  \phi_{n+\hat{\mu}} + \ln \det \,[\dsl +  y
\phi] + \sum_n J_n \phi_n}.}  Then for free energy of the system $F = W[0]$
the
method yields the inequality \eqn\MF{F \leq F_{MF} = \inf_{h_n} [-\sum_n
(u(h_n)
- h_n u'(h_n)) - \vev{ 2 \kappa \sum_{n, \mu} \phi_n  \phi_{n+\hat{\mu}} +
\ln
\det \,[\dsl +  y \phi]}_h ],}
where $h_n$ is a mean field, and
\eqn\u{\eqalign{& u(h_n) = \ln \sum_{\phi_n \in Z_2} \e{h_n \phi_n} = \ln 2
\cosh h_n, \cr & \vev{O[\phi]}_h = \e{-\sum_n u(h_n)}
\sum_{\phi_n \in Z_2} O[\phi] \e{\sum_n h_n \phi_n}.\cr}}
So, we can get some idea of the system, studying $F_{MF}$, that is much
simpler
than that for $F$. From \u\ it immediately follows that
\eqn\corr{\eqalign{& \vev{\phi_n} = u'(h_n) = \tanh h_n, \cr
&\vev{\phi_m\,\phi_n}_h = u'(h_m) u'(h_n) + \delta_{mn} u''(h_m), \;\;\;
etc.,
\cr}}
and therefore the main problem is a calculation of the expectation value of
the fermionic determinant
\eqn\d{\eqalign{\vev{\ln &\det \,[\dsl +  y \phi]}_h \cr
& = \ln \det [\dsl]  -
\sum^{\infty}_{n = 1} {(-1)^{n}\over n} y^n \sum_{i_1,...,i_n} \tr\,
(\dsl^{-1}_{i_1\,i_2}\dsl^{-1}_{i_2\,i_3}...\dsl^{-1}_{i_n\,i_1})
\vev{\phi_{i_1}\phi_{i_2}...\phi_{i_n}}_h \cr & = 2^{D/2} N^D \ln y  -
\sum^{\infty}_{n = 1} {(-1)^{n}\over n} {1\over y^n} \sum_{i_1,...,i_n} \tr\,
(\dsl_{i_1\,i_2}\dsl_{i_2\,i_3}...\dsl_{i_n\,i_1})
\vev{\phi_{i_1}\phi_{i_2}...\phi_{i_n}}_h, \cr}}
where $\tr$ stands for the trace over spinorial indices; in the first term of
the second equation the relation $\phi_i^{2^{D/2}} = 1$ has been taken into
account.

Following the usual way we consider $F_{MF}$ for two translation invariant
ansatzes for $h_n$
\eqn\h{\eqalign{& h_n^{FM} = h, \cr & h_n^{AF} = \epsilon_n h,\;\;\;
\epsilon_n =
(-1)^{\sum_{\mu} n_{\mu}}.\cr}}
which in fact are the order parameters distinguishing the ferromagnetic (FM:
$h_n^{FM} \neq 0$, $h_n^{AF} = 0$),  antiferromagnetic (AF: $h_n^{FM} = 0$,
$h_n^{AF} \neq 0$), paramagnetic (PM: both are zero), and ferrimagnetic (FI:
both are nonzero) phases in the system. Then the mean field equations are
reduced to
\eqn\mfe{{\del \over \del h} F_{MF}^{FM,\,AF} = 0,}
where $F_{MF}^{FM,\,AF}$ is the functional of the right-hand side of Eq.(3.2)
on ansatzes (3.6).
Further simplification comes from the observation (see, for example \STs),
that
as the value $h = 0$ is always a solution of Eq. (3.7), and, therefore,
second
order phase transition lines are determined by equations
\eqn\tl{{\del^2 \over \del h^2} F_{MF}^{FM,\,AF}\mid_{h = 0} = 0,}
to find them it is sufficient to know $\vev{\ln \det \,[\dsl +  y \phi]}_h$
to
terms of order of $h^2$.

If the problem could be solved exactly both of two representations (3.5) of
the
fermionic determinant would yield the same answer. But correlations of
$\phi{^,}$s at coinciding arguments (Eq.(3.4)) make the problem unsolvable
exactly, as the contributions of order of $h^2$ to (3.5) come from terms of
any
orders of $u''$, as well as from those of order of $u'^2$. These
contributions
shown schematically in Fig. 1. Therefore, we are forced to use some
approximations, and, particularly, to use two representations of (3.5)
separately for ``weak" and ``strong" coupling regimes of $y$, though the
exact
meaning of this can only be clear a posteriori.

Our approximation involves summing up all diagrams of Fig.1 ($a$) (proper
ladder diagrams) and ($b$) (crossed ladder diagrams), so  we may call it
as a ladder approximation. Using property (2.4) of the Dirac operator we find
that the contributions to $F_{MF}^{FM, AF}$ from the  fermionic determinant,
$\Delta F_{MF}^{FM, AF}$, have the same functional form for both
representations
(3.5) and in our approximation read as follows
\eqn\cfd{\eqalign{&\Delta F_{MF}^{FM} = N^D 2^{D/2 - 1} ({c^2
u'^2 G(0) \over 1 + c^2 u'' G(0)} + N^{-D} \sum_q {c^2 u'' G(q)
\over 1 + c^2 u'' G(q)}) \cr
& \Delta F_{MF}^{AM} = N^D 2^{D/2 - 1} ({c^2 u'^2 G(\pi)
\over 1 + c^2 u'' G(\pi)} + N^{-D} \sum_q {c^2 u'' G(q)
\over 1 + c^2 u'' G(q)}), \cr}}
where $q_{\mu} = (2 \pi /N) l_{\mu}$, $-N/2+1 \leq l_{\mu}\leq N/2$ (so that
$q_{\mu} \in (-\pi, \pi])$, while coupling $c$ and form of function $G$
depend
on the representation. So, for weak coupling regime we have $c = y$ and
\eqn\gw{G^W (q) =  N^{-D} \sum_p {L(p) L(p+q) \over L^2 (p) L^2(p+q)} =
N^{-D}
\sum_p{\sum_{\mu} \sin (p)_{\mu} \sin (p+q)_{\mu} \over \sum_{\mu}   \sin^2
p_{\mu} \sum_{\nu} \sin^2 (p+q)_{\nu} },}   and for the strong coupling they
are
$c = y^{-1}$ and \eqn\gs{ G^S (q) = {1\over N^D} \sum_p L(p) L(p+q)  =
N^{-D} \sum_{p,\mu} \sin p_{\mu} \sin (p+q)_{\mu}.}
The first terms in (3.9) come from the diagrams of Fig.1($a$), while the
second from those of Fig.1($b$).

Then, from Eq. (3.8) and the above formulae it follows
that critical lines in the system in our approximation are determined by the
expressions   \eqn\res{\eqalign{
& \kappa^{F(W)}_{cr} = {1 \over 4D} [ 1 - 2^{D/2} y^2 ( { G^W(0) \over 1 +
y^2
G^W (0)} - N^{-D} \sum_q  { G^W (q) \over (1 + y^2 G^W (q))^2})], \cr
&\kappa^{AF(W)}_{cr} = -{1 \over 4D} [ 1 - 2^{D/2} y^2 ( { G^W (\pi)
\over 1 + y^2 G^W (\pi)} - N^{-D} \sum_q  { G^W (q) \over (1 + y^2 G^W
(q))^2})];\cr
&\kappa^{F(S)}_{cr} = {1 \over 4D} [ 1 - 2^{D/2} ( { G^S (0) \over y^2 +
G^S (0}
- N^{-D} \sum_q  { y^2 G^S (q) \over (y^2 + G^S (q))^2})], \cr
&\kappa^{AF(S)}_{cr}
= -{1 \over 4D} [ 1 - 2^{D/2} ( { G^S (\pi) \over y^2 + G^S (\pi)} - N^{-D}
\sum_q  { y^2 G^S (q) \over (y^2 + G^S (q))^2})]. \cr }}

We now should make some comments.

(i) The contributions to (3.12) which are
proportional to $G(0)$ and $G(\pi)$ are generalization of
"double chain" contributions of Ref. \STs, as the diagrams of Fig.1($a$) are
the generalization of the double chains to any configurations of the same
topology. They coincide only for $G^S$ because of strict locality of
the Dirac operator, but not for $G^W$. More important difference comes from
the second terms corresponding to the diagrams of Fig.1($b$) (the latter
correspond to the generalization of the double chains with coinciding ends),
which have not been taken into account in previous calculations (see also
\ref\EK{T. Ebihara, K.-I. Kondo, Nucl. Phys. B (Proc. Suppl.) 26 (1992)
519}).
{}From the well known symmetry of the model under the transformations:
$(\psi, \psibar)_n \rightarrow \exp (i \epsilon_n \pi /4) (\psi, \psibar)_n$,
$\phi_n \rightarrow \epsilon_n \phi_n$, $\kappa \rightarrow -\kappa$, $y
\rightarrow -i y$, it follows that $G(\pi) = -G(0)$, and also, that
the contributions of the new terms are of even power in $y^{\pm 2}$ beginning
from $y^{\pm 4}$.

(ii) These terms can become dominating when $y^2$ is close to the values
$1/G^W(0)$ or $G^S(0)$ which are singular points of the expressions under the
sum, even though in weak coupling regime they are of $O(D^{-1})$ compared
with
the first ones. Thereby these terms determine domains of the ``weak" and
``strong" coupling regimes also for $\kappa^F_{cr}$. They are domains of
analyticity of functions $\kappa^W_{cr}(y)$ and $\kappa^S_{cr}(y)$, that is
$y^2 < 1/G^W (0)$ and  $y^2 > G^S (0)$, respectively, coinciding for
$\kappa^F_{cr}$ and $\kappa^{AF}_{cr}$.

(iii) We have no strict arguments why other diagrams which we did not take
into
account could be neglected compared with the ladder ones. In particular, in
strong coupling regime they can give contributions to $\kappa_{cr}$ of the
same
order in $1/D$ as the latter. But because those diagrams come into play in
higher orders in $y^{\pm 1}$, at least from the order of $y^{\pm 6}$, the
assumption that their contributions are suppressed and less singular looks
plausible.

Finally, it worth noting that the formulae (3.12) are applicable to any
lattice fermion actions, including non-local ones, whose Dirac operators
satisfy
property (2.4) \ref\Z{S. V. Zenkin, "The phase structure of
Higgs-Yukawa models with chirally invariant lattice fermion actions",
preprint
KYUSHU-HET-8 (1993)}.

\newsec{Results and discussion}

Let us now compare the phase diagrams determined by the expressions (3.12)
for
$D=4$ for finite $N$ and for the limiting case of $N \rightarrow \infty$ .
The new terms are always negative and therefore increase the contributions of
the first terms for $\kappa^F_{cr}$ and decrease them for $\kappa^{AF}_{cr}$.
The question is of how much.

For $N = 4$ we have $G^W(0) = 0.5$, $G^S(0) = 2$, so that the domain of
inapplicability of our formulae shrinks to the point $y = 2^{1/2}$, and the
phase diagram is shown in Fig. 2. The curves $\kappa^F_{cr}(y)$ and
$\kappa^{AF}_{cr}(y)$ intersect each other forming narrow domain with the
ferrimagnetic phase around $y = 2^{1/2}$, which is spreaded from $-\infty$ to
$\infty$ in $\kappa$. It is natural to assume, that contributions of
other diagrams (Fig.1($c$)) smooth the negative contribution of those of Fig.
1($b$), so that the PM-AF phase transition line in Fig. 2($b$) becomes
continuous. Then, as a result we would have a familiar picture, typical for
SU(2) models (see, for example, \ref\Bea{W.
Bock, A. K. De, K. Jansen, J. Jers\'{a}k, T. Neuhaus, J. Smit, \NP{344}
(1990)
207}), with FI phase lying below this line.

In the limit of $N \rightarrow \infty$ we have $G^W(0) \simeq 0.62$, $G^S(0)
=
2$, but the picture is changed qualitatively. The phase diagram is shown in
Fig. 3. The curves do not touch each other even at $\kappa \rightarrow
-\infty$
and FI phase does not appear.

To clear up why this happens let us consider behaviour of functions of $y^2$
determined by the sums in (3.12) near the points $1/G^W(0)$ and $G^S(0)$.
Let us define positive $\delta = y^2 - G^S(0)$ or $1/G^W(0) - y^2$. Then a
simple analysis shows that at a finite $N$ and a small $\delta^,$s these
functions is of order of $O(D^k N^{-D} \delta^{-2})$, so the intersections of
the curves $\kappa^F_{cr}(y)$ and $\kappa^{AF}_{cr}(y)$ always occur at the
points $\delta = O(D^l N^{-D})$, $-\kappa = O(D^m 2^{D/2}  N^D)$, where $k$,
$l$, $m$ are some (negative or non-negative) powers. But at $N \rightarrow
\infty$, when the sums go over to integrals, this functions become of order
of
$\ln \delta$ for $D = 4$, and even of $O(\delta^0)$ at $D > 4$. This means
that
at $D > 4$ we can continue lines $\kappa^{F(W)}_{cr}(y)$ and
$\kappa^{F(S)}_{cr}(y)$ until they intersect each other, so that the phase
diagram in this case looks like in Fig. 4, that reproduces the result of
ref.\STs.

Thus, this example demonstrates importance of summing up contributions to
fermionic determinant including those of the next order in $1/D$ for $D = 4$
systems. Another point is that even though we cannot definitely conclude
whether
the FI phase in this example is an artifact only of a finite lattice or also
of
the mean field approximation, this gives one one more caution in what
concerns
finite lattice effects.

\bigbreak\bigskip
\centerline{{\bf Acknowledgments}}
\vskip 8pt

This work is supported by JSPS. I am grateful to H. Yoneyama for enlightening
discussions and careful reading the manuscript and to S. Sakoda for drawing
the
picture of Fig. 1. It is also a pleasure to thank all members of the
Elementary
Particle Theory Group of the Department of Physics of Kyushu University for
their warm hospitality.
\vfil\eject
\listrefs
\centerline{}

\vskip 50ex

\epsfxsize=\hsbody
\multiply\epsfxsize by 8
\divide\epsfxsize by 10
\epsfbox{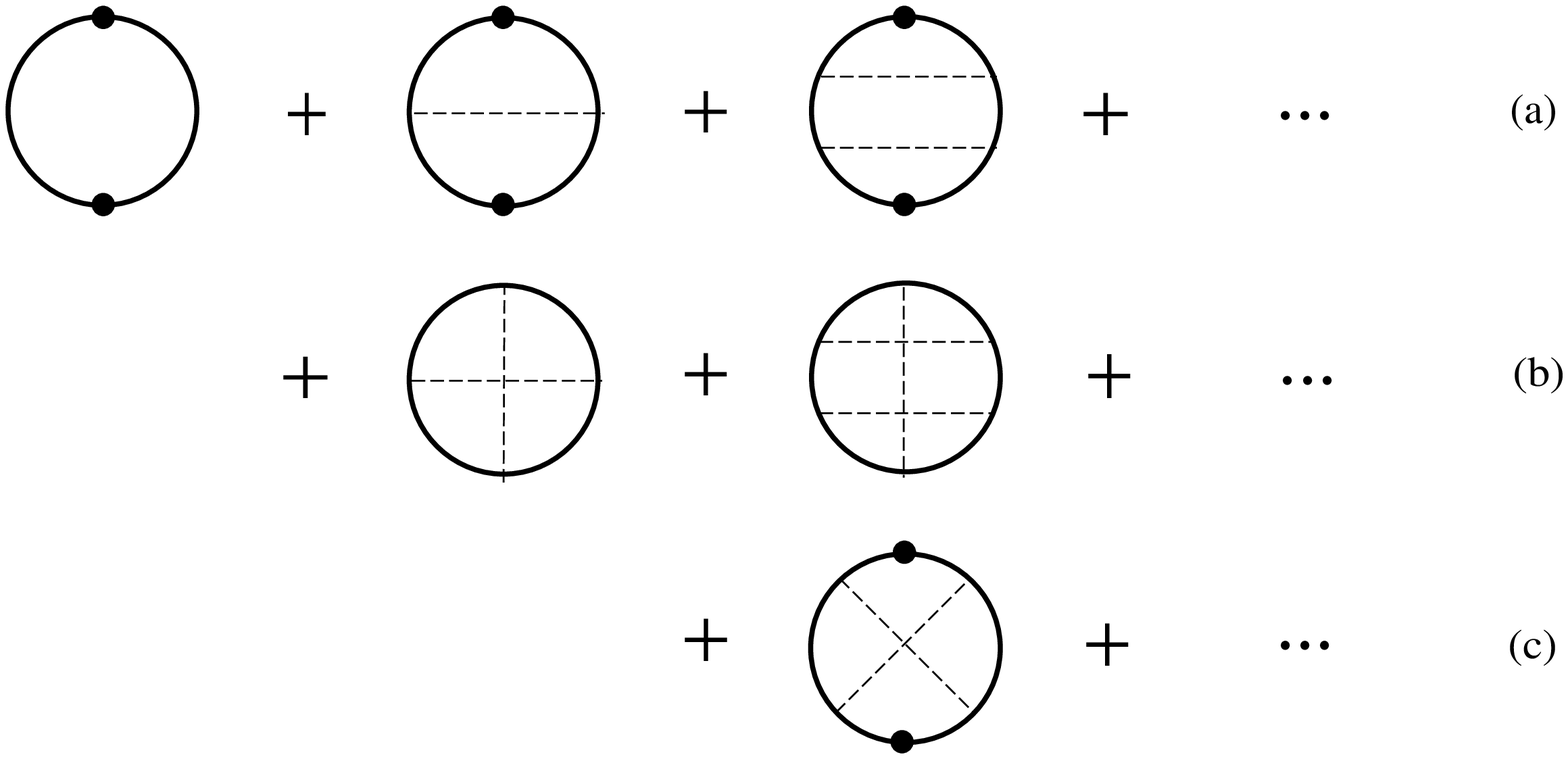}
\bigskip
Fig. 1. Diagrams contributed to $\Delta F$ to order
$h^2$. Solid lines denote $\dsl$ or $\dsl^{-1}$, each solid circle stands for
$u'$, dashed line for $u''$.
\vfil\eject
\centerline{}

\vskip 10ex

\epsfxsize=\hsbody
\multiply\epsfxsize by 8
\divide\epsfxsize by 10
\epsfbox{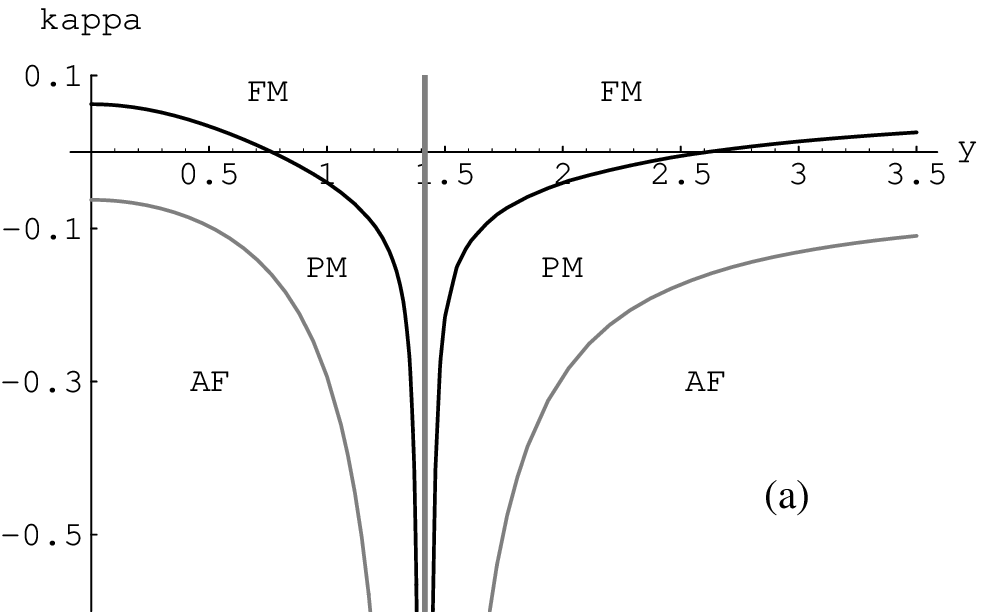}
\smallskip
\epsfxsize=\hsbody
\multiply\epsfxsize by 8
\divide\epsfxsize by 10
\epsfbox{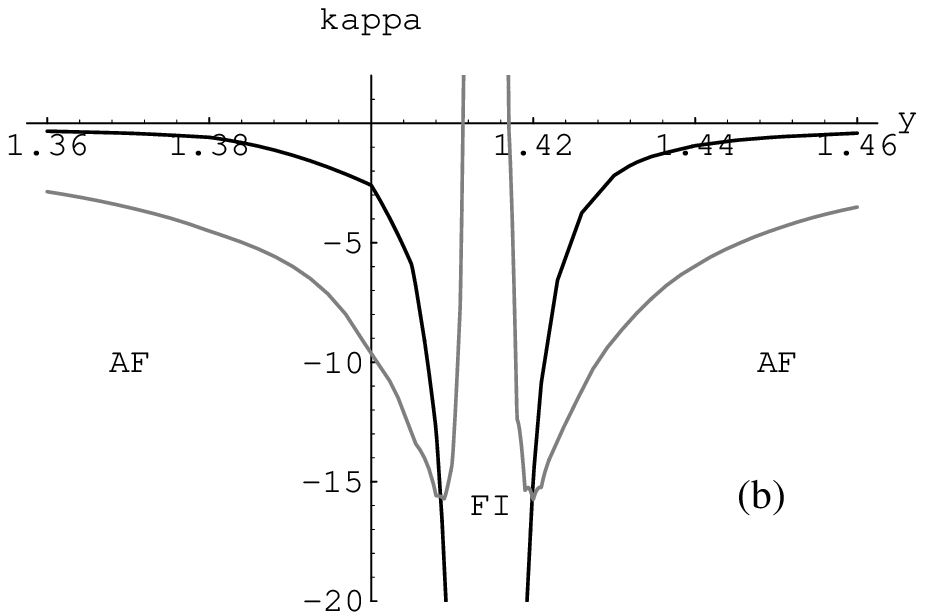}
\smallskip
Fig. 2. $(a)$ Phase diagram of the model
at $D = 4$, $N = 4$. Intersections of FM-PM phase transition line (solid)
with
PM-AF phase transition line (gray) form FI phase in the narrow region around
the
point $y = 2^{1/2}$  shown in $(b)$ in more detail.
\vfil\eject
\centerline{}

\vskip 10ex

\epsfxsize=\hsbody
\multiply\epsfxsize by 8
\divide\epsfxsize by 10
\epsfbox{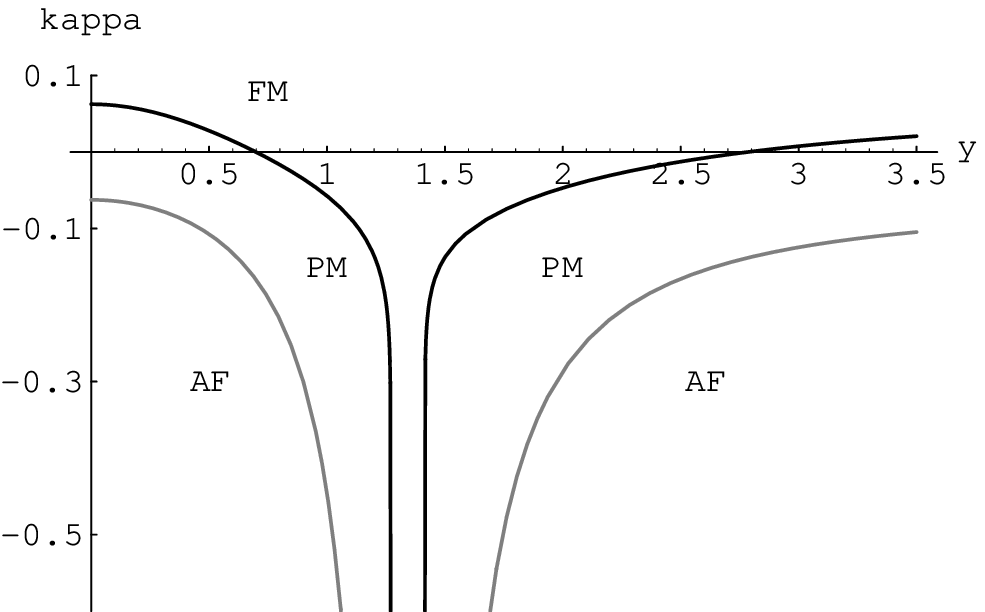}
\smallskip
Fig. 3. Phase diagram of the model at $D = 4$, $N
\rightarrow \infty$.
\smallskip
\epsfxsize=\hsbody
\multiply\epsfxsize by 8
\divide\epsfxsize by 10
\epsfbox{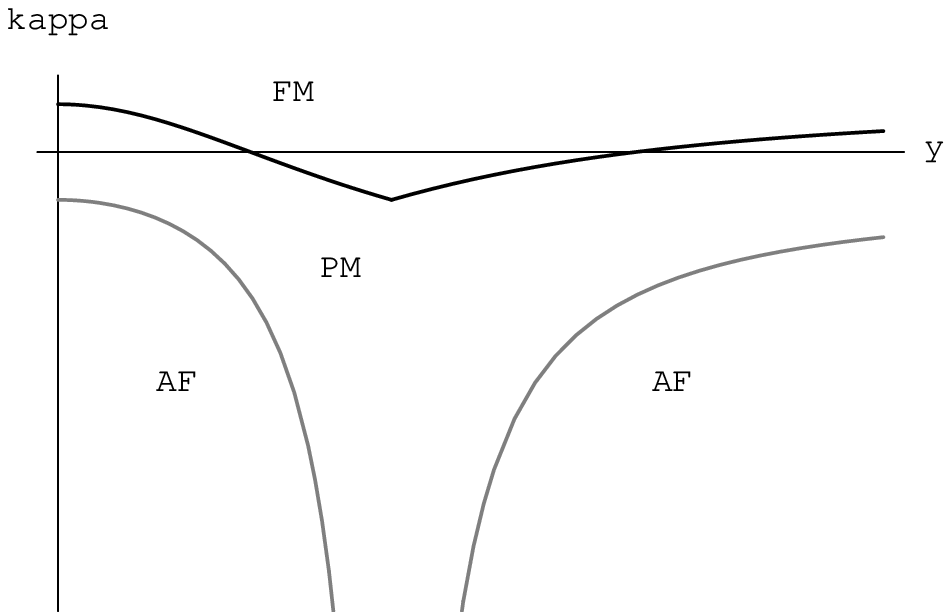}
\smallskip
Fig. 4. Qualitative picture of the phase diagram of the
model at $D > 4$, $N \rightarrow \infty$.
\bye